\documentclass[12pt]{article}
\usepackage{graphicx}

\newcommand\pubnumber{DPF2015-371}
\newcommand\pubdate{\today}

\def\msu{Department of Physics and Astronomy\\
          Michigan State University, East Lansing, MI 48824, USA
          }
\def\support{\footnote{This contribution, presented by Pawin Ittisamai, is an abbreviated version of \cite{Chivukula:2015zma}.}}
\def\currentaddress{\footnote{Current address: Department of Physics, Faculty of Science, Chulalongkorn University, Bangkok 10330, Thailand. Email: \textit{ittisama@msu.edu}}}

\def\Title#1{\begin{center} {\Large #1 } \end{center}}
\def\Author#1{\begin{center}{ \sc #1} \end{center}}
\def\Address#1{\begin{center}{ \it #1} \end{center}}

\newcommand\pubblock{\rightline{\begin{tabular}{l} \pubnumber\\
         \pubdate  \end{tabular}}}
\newenvironment{Abstract}{\begin{quotation}  }{\end{quotation}}
\newenvironment{Presented}{\begin{quotation} \begin{center}
             PRESENTED AT\end{center}\bigskip
      \begin{center}\begin{large}}{\end{large}\end{center} \end{quotation}}
\def\Acknowledgments{\bigskip  \bigskip \begin{center} \begin{large}
             \bf ACKNOWLEDGMENTS \end{large}\end{center}}
\def\beq{\begin{equation}}
\def\eeq{\end{equation}}
\def\bea{\begin{eqnarray}}
\def\eea{\end{eqnarray}}
\def\bealn{\begin{eqnarray}}
\def\eealn{\end{eqnarray}}

\def\ifb{\rm fb^{-1}}

\def\tev{\rm TeV}
\def\zp{{Z^\prime}}	
\def\dcol{D_{\rm{col}}}

\begin{document}
\begin{titlepage}
\pubblock

\vfill
\Title{Color Discriminant Variable to Separate Dijet Resonances at the LHC}
\vfill
\Author{Pawin Ittisamai\support\currentaddress, R. Sekhar Chivukula, Kirtimaan Mohan, and Elizabeth H. Simmons}
\Address{\msu}
\vfill
\begin{Abstract}
A narrow resonance decaying to dijets could be discovered at the 14 TeV run of the LHC. To quickly identify its color structure in a model-independent manner, we introduced a method based on a color discriminant variable, determined from the measurements of the resonance's production cross section, mass and width. This talk introduces a more transparent theoretical formulation of the color discriminant variable that highlights its relationship to the branching ratios of the resonance into incoming and outgoing partons and to the properties of those partons. The formulation makes it easier to predict the value of the variable for a given class of resonance. We show that this method applies well to color-triplet and color-sextet scalar diquarks, distinguishing them clearly from other candidate resonances.

\end{Abstract}
\vfill
\begin{Presented}
DPF 2015\\
The Meeting of the American Physical Society\\
Division of Particles and Fields\\
Ann Arbor, Michigan, August 4--8, 2015\\
\end{Presented}
\vfill
\end{titlepage}
\def\thefootnote{\fnsymbol{footnote}}
\setcounter{footnote}{0}

\section{Introduction}

\label{sec:intro}
A scalar or vector resonance coupled to quarks in the standard model can be abundantly produced at a hadron collider of sufficient energy. Then it decays to a final state of simple topology: dijet (a pair of jet), top quarks, or bottom quarks, both of which are highly energetic and clustered in the central region of the detector. In a large data sample, a resonance with a relatively small width will appear as a distinct bump over a large, but exponentially falling, QCD background. These features make the hadronic decay channels favorable for discovery.

There have been numerous searches for Beyond-the-Standard-Model (BSM) resonances decaying to dijet final states at colliders. As no new dijet resonances have been discovered so far, the current exclusion limits on the production cross section for those of sufficiently narrow width have been set by searches carried out by ATLAS and CMS collaborations at the LHC with a center-of-mass energy of $8 \,\tev$~\cite{ATLAS:2012pu, Chatrchyan:2013qha, CMS:kxa}. The upgraded, higher-energy LHC will be able to seek a resonance with a larger mass, and the greater integrated luminosity will enable the experiments to reach new discovery thresholds.

When the LHC discovers a new dijet resonance, it will be crucial to determine the spin, color, and other properties of the resonance in order to understand what kind of BSM context it represents. In previous work~\cite{Atre:2013mja}, we introduced a way to distinguish whether a vector resonance is either a leptophobic color-singlet or a color-octet, using a construct that we called a ``color discriminant variable'', $\dcol$. The variable is constructed from the dijet cross-section for the resonance ($\sigma_{jj}$), its mass ($M$), and its total decay width ($\Gamma$), observables that will be available from the dijet channel measurements of the resonance:
	\beq
		\dcol \equiv \frac{M^3}{\Gamma} \sigma_{jj}.
	\label{eq:dcol}
	\eeq
For a narrow-width resonance, the color discriminant variable is independent of the resonance's overall coupling strength.

We have applied the color discriminant variable technique both to flavor universal vector resonances with  identical couplings to all quarks~\cite{Atre:2013mja} and also to more generic flavor non-universal vector resonances~\cite{Chivukula:2014npa} whose couplings to quarks vary by electric charge, chirality, or generation.  In the latter case, combining the color discriminant variable with  information from resonance decays to heavy top ($t\bar{t}$) or bottom ($b\bar{b}$) flavors still enables one to determine what type of resonance has been discovered.  We have also shown~\cite{Chivukula:2014pma} that the method can be used to separate fermionic or scalar dijet resonances from vector states.

In this talk, we present the extension of the color discriminant variable technique in two directions. First, we re-frame the theoretical discussion of the variable in more general language that shows its broader applicability and its relationship to the properties of the partons involved in production and decay of a narrow resonance.  In addition, we show that $D_{col}$ can be used to distinguish a color-triplet or color-sextet scalar diquark (a weak-singlet state coupling to two quarks) from weak-singlet vector dijet resonances that couple to a quark/anti-quark pair, such as a coloron (color-octet) or $Z'$ (color-singlet).

\section{The Color Discriminant Variable}
\label{sec:coldis}
Searches for new particles currently being conducted at the LHC are focused on resonances having a narrow width. So one can expect that if a new dijet resonance is discovered, the  dijet cross section, mass,  and width of the resonance will be measured. These three observables are exactly what is needed to construct the color discriminant variable, as defined in (\ref{eq:dcol}) that can distinguish between resonances of differing color charges and spin states.

We present a formulation of the tree-level $s$-channel resonance cross section which makes the properties of the color-discriminant variable more transparent and makes $D_{col}$ easier to calculate for diverse types of resonances. Following Eq.~(44) of \cite{Harris:2011bh}, the spin- and color-averaged partonic tree-level $s$-channel cross section for the process $i + k \to R \to x + y$ is written
\begin{equation}
\hat{\sigma}_{ik\to R\to xy}(\hat{s}) = 16 \pi \cdot {\cal N} \cdot (1 + \delta_{ik}) \cdot
\frac{\Gamma(R\to ik) \cdot \Gamma(R\to xy)}
{(\hat{s}-m^2_R)^2 + m^2_R \Gamma^2_R} ~,
\end{equation}
where $(1 + \delta_{ik})$ accounts for the possibility of identical incoming partons.  The factor ${\cal N}$ is a ratio of spin and color counting factors
\begin{equation}
{\cal N} = \frac{N_{S_R}}{N_{S_i} N_{S_k}} \cdot
\frac{C_R}{C_i C_k}~,
\end{equation}
where $N_S$ and $C$ count the number of spin- and color-states for initial state partons $i$ and $k$. Using the narrow-width approximation, integrating over parton densities, and summing over incoming partons, as well as the outgoing partons that produce hadronic jets ($jj$), we then find the tree-level hadronic cross section to be
\begin{equation}
\sigma_R =
16\pi^2 \cdot {\cal N} \cdot \frac{ \Gamma_R}{m_R} \cdot
\left(\sum_{xy = jj} BR(R\to xy)\right)
\left( \sum_{ik} (1 + \delta_{ik}) BR(R\to ik) \left[\frac{1}{s} \frac{d L^{ik}}{d\tau}\right]_{\tau = \frac{m^2_R}{s}}\right)~,
\end{equation}
where $ d { L}^{ud}/ d\tau$ is the parton luminosity function. Hence, for the color discriminant variable defined in Eq.~(\ref{eq:dcol}), we find the general expression
\begin{equation}
D_{col}=  16\pi^2 \cdot {\cal N}  \cdot
\left(\sum_{xy=jj} BR(R\to xy)\right)
\left( \sum_{ik} (1 + \delta_{ik}) BR(R\to ik) \left[\tau \frac{d L^{ik}}{d\tau}\right]_{\tau = \frac{m^2_R}{s}}\right)~.
\label{eq:dcol-expression}
\end{equation}
This expression illustrates the dependence of the color discriminant variable on the properties of the incoming and outgoing partons, and can easily be applied to any narrow resonance.

Here, as an example, we consider a color-triplet weak-singlet scalar diquark $\omega_3$, produced via collisions of $u$ and $d$ quarks and decays back to the same quark pair:  $pp \to u_L d_L \to \omega_3 \to u_L d_L$. The diquark has $N_{S_R} =1 $ and $C_R = 3$; each incoming quark has $N_{S_i} = N_{S_k} = 2$ and $C_i = C_k = 3$; and there is only one incoming and one outgoing mode, we find:
\beq
D_{col}^{\omega_3} =  \frac{4\pi^2}{3}
\left[ \tau \frac{d { L}^{ud}}{d\tau} \right]_{\tau = \frac{m^2_{\omega_3}}{s}}~.
\label{eq:dcol-expression-D-alt}
\eeq

\section{Accessibility at the 14 TeV LHC}
\label{sec:paramspace}

After a narrow dijet resonance has been discovered, one uses the measurements of three observables; dijet cross section, mass, and total decay width to evaluate $\dcol$ via Eq.~(\ref{eq:dcol}).  At the same time, one can use Eq.~(\ref{eq:dcol-expression}) to compare $\dcol$ with the predictions for various classes of dijet resonances.

\begin{figure}[htb!]
\centering
{
\includegraphics[width=0.49\textwidth, clip=true]{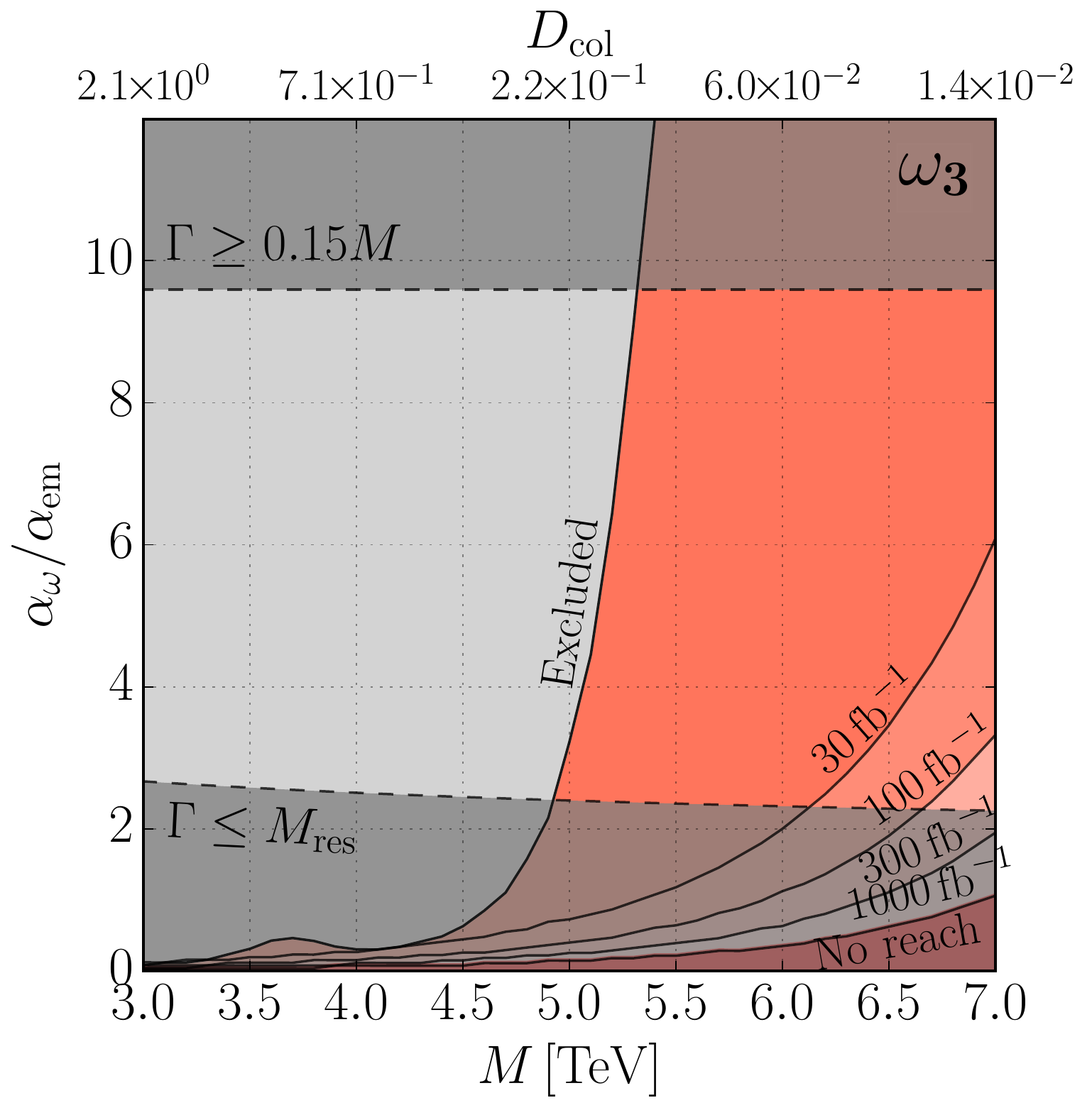}
}
\caption{\small
Viable parameter (here $\alpha_\omega = \lambda^2/4\pi$, and is plotted relative to the electromagnetic coupling $\alpha_{em}$) space, the salmon-shaded area to the right of the curve labeled ``excluded", for the color triplet diquark $\omega_3$, which couples to $u_L d_L$. The curve at lower right labeled ``$30\, \ifb$'' delimits the region accessible to a $5\sigma$ discovery at the 14 TeV LHC with that integrated luminosity; the curves below it show how higher integrated luminosities (respectively, from above, 100, 300, 1000 $\ifb$) would increase the reach.  The region in which $D_{col}$ can be measured lies below dashed curve line where the resonance width equals 15\% of its mass and above the dashed curve where the resonance width equals the mass resolution of the detector is amenable; areas where the resonance is too broad or too narrow have been given a cloudy overlay.  The lowest red-shaded region lies beyond the reach of 1000 $\ifb$ of data.}
\label{fig:param_space_triplet}
\end{figure}

We describe the region of parameter space to which the method is applicable. In this region the resonance has not already been excluded by the current searches, can be discovered at the $5\sigma$ level at the LHC 14 TeV after statistical and systematic uncertainties are taken into account, and has a total width that is measurable and consistent with the designation ``narrow''. Fig.~\ref{fig:param_space_triplet} shows the viable region of parameter space for a color triplet diquark%
\footnote{
	Following the notation in~\cite{Han:2009ya}, we write the interactions of the diquark states $\omega_3$ and $\tilde{\omega}_3 \equiv u_R d_R$ with quarks as
$$
 {\cal L} = 2\sqrt{2}\left(\bar{K}_3\right)^{ab}_c\left[
 \lambda_\omega \omega_3^c \bar{u}_{La} d^C_{Rb} + \lambda_{\tilde{\omega}} \tilde{\omega}_3^c \bar{u}_{Ra} d^C_{Lb}\right]+h.c.~,
$$
where $a,b$ and $c$ are color (triplet) indices, $\bar{K}_3$ is the color Clebsch-Gordan coefficient connecting two triplets to an anti-triplet (related to $\epsilon_{abc}$), and $\lambda_{\omega,\tilde{\omega}}$ are unknown coupling constants.
}
$\omega_3$.

Statistical and systematic uncertainties on dijet cross section, mass, and intrinsic width of the resonance will play a key role in determining how well $\dcol$ can discriminate between models at the LHC with $\sqrt{s} = 14\,\tev$.  While the actual values of the systematic uncertainties at the LHC with $\sqrt{s} = 14\,\tev$ will be obtained only after the experiment has begun, we have previously discussed estimates of the uncertainties in \cite{Atre:2013mja,Chivukula:2014npa,Chivukula:2014pma}.  In particular, we reviewed estimates of the effect of systematic uncertainties in the jet energy scale, jet energy resolution, radiation and low mass resonance tail and luminosity on the dijet cross section at the 14 TeV LHC from  Ref.~\cite{Gumus:2006mxa} and discussed how this, combined with the dijet mass resolution would impact measurements of $\dcol$. Overall, it appears that uncertainties of 20 - 50\% should be achievable and with that level of accuracy the color discriminant variable should be a useful tool for distinguishing among dijet resonances.


\section{Distinguishing Diquarks from Vector Bosons}
\label{sec:result-dcols}
We illustrate how the color discriminant variable $\dcol$  may be used to distinguish whether a newly discovered dijet resonance is a scalar diquark, as opposed to a coloron or a leptophobic $\zp$.  We will focus on resonances having masses of $3-7\,\tev$ at the $\sqrt{s} = 14\,\tev$ LHC with integrated luminosities up to $1000\,\ifb$. The values of $\dcol$ and other observables have been evaluated using the uncertainties discussed in \cite{Atre:2013mja,Chivukula:2014npa,Chivukula:2014pma} and the region of parameter space to which this analysis is applicable was identified in Section \ref{sec:paramspace}.

\begin{figure}[htb!]
{
\includegraphics[width=0.49\textwidth, clip=true]{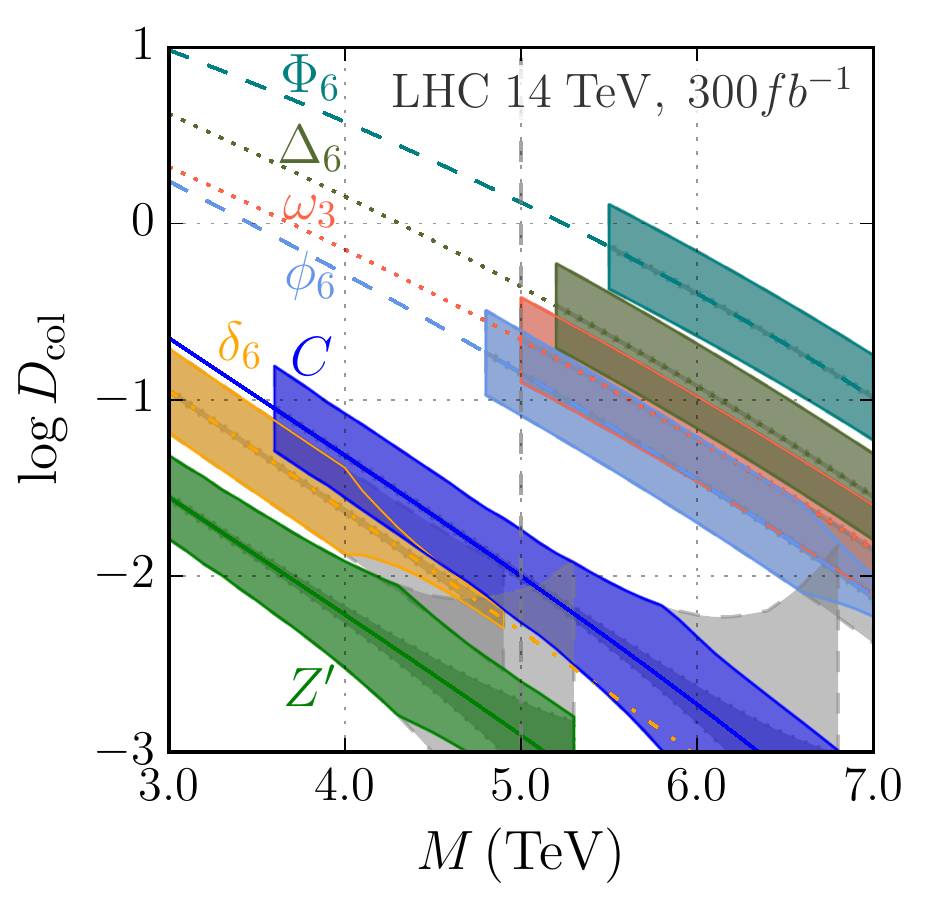}
\includegraphics[width=0.49\textwidth, clip=true]{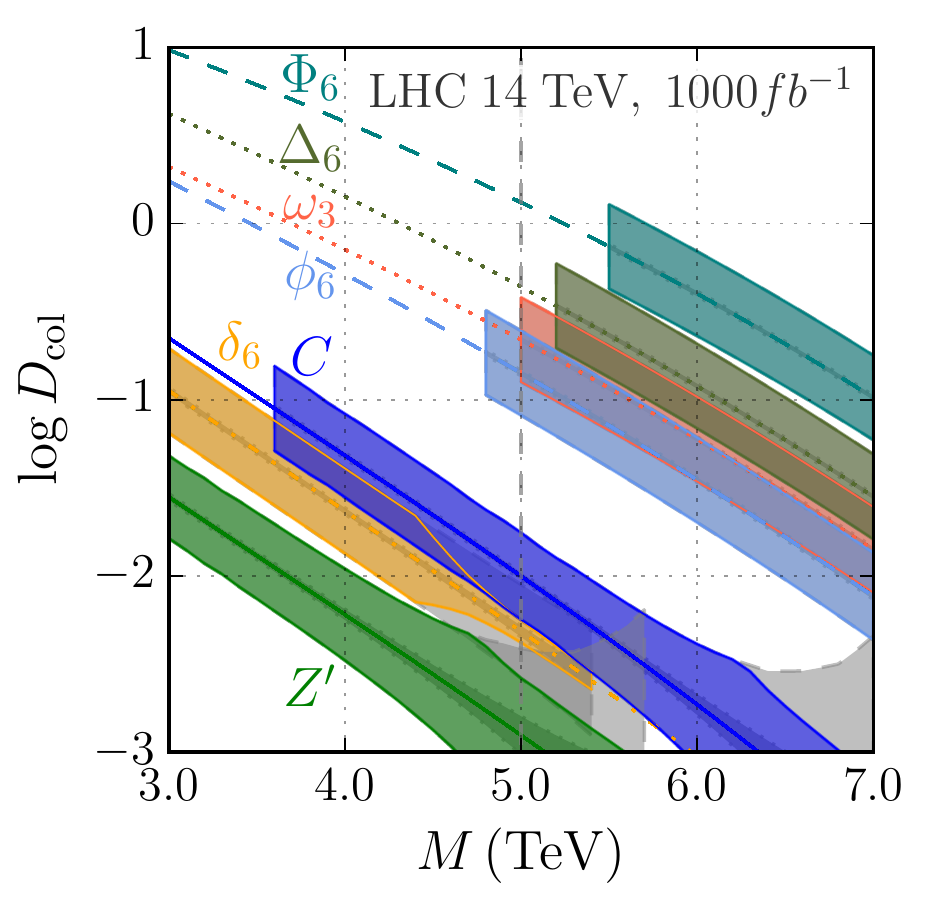}
}
\caption{\small
Color discriminant variables calculated, as described in Sec.~\ref{sec:paramspace}, using estimated systematic and statistical uncertainties for mass, total width, and dijet cross section for the integrated luminosities $300\,~\ifb$ (Left) and $1000\,~\ifb$ (Right).
The central value of $\dcol$ for each particle is shown as, from top to bottom: $\Phi_6$ (dashed red), $\Delta_6$ (dotted black), $\omega_3$ (dotted red), $\phi_6$ (dashed black), $C$ (solid blue), $\delta_6$ (dotted-dash yellow), $\zp$ (solid green).
The uncertainty in the measurement of $\dcol$ due to the uncertainties in the measurement of the cross section, mass and width of the resonance is indicated by gray bands.
The outer (darker gray) band corresponds to the uncertainty in $\dcol$ when the width is equal to the experimental mass resolution i.e. $\Gamma = M_{\mathrm res}$. The inner (lighter gray) band corresponds to the case where the width $\Gamma= 0.15M$. Resonances with width $M_\mathrm{res} \leq \Gamma \leq 0.15M$ will have bands whose widths fall between the outer and inner gray bands.
The horizontal extent of the colored band for each state indicates the region in parameter space where the particle has not been excluded by current searches and has the potential to be discovered at the $5\sigma$ level at the LHC 14 TeV after statistical and systematic uncertainties are taken into account.
The limits and reaches for resonances with masses above $5\,\tev$ have been extrapolated; as a reminder, the extrapolated region lies to the right of a vertical dashed line.
}
\label{fig:pcombined-dcols-sys-stat-errors_300_1000}
\end{figure}

Figure~\ref{fig:pcombined-dcols-sys-stat-errors_300_1000} compares the value of $D_{col}$ as a function of resonance mass for several different resonances: colorons, $\zp$ bosons, the color-triplet diquark%
\footnote{
	The diquark states are $\omega_3 \equiv u_L d_L$, $\Phi_6 \equiv u_R uR$, $\phi_6 \equiv d_R d_R$, $\Delta_6 \equiv u_R d_R$, and $\delta_6 \equiv (u_L s_L - c_L d_L)$.
	See Ref.~\cite{Chivukula:2015zma} for the complete descriptions of the diquarks studied in this work.
	}
$\omega_3$ and the color-sextet diquarks $\Phi_6$, $\phi_6$, $\Delta_6$ and $\delta_6$; for 300 and 1000 $\ifb$.  In each plot, a given colored band shows the mass range in which the corresponding resonance is viable and accessible.  The appropriate exclusion limit in Fig.~\ref{fig:param_space_triplet} delimits the left-hand edge of each band; the appropriate integrated luminosity curve from Fig.~\ref{fig:param_space_triplet} delimits the right-hand edge, beyond which there is not enough data to allow discovery at a given mass.  The width of each band relates to measurement uncertainties, as detailed in the figure caption.

For a given value of the dijet resonance mass, some types of resonance may already be excluded (e.g., a resonance found below about 3.4 TeV cannot be a coloron), while others may lie beyond the LHC's discovery reach at a given integrated luminosity, because too few events would be produced.  But for any resonance mass between about 3.5 and 7 TeV, there are generally several dijet resonances that remain viable candidates For example, a resonance discovered at 4.0 TeV could be a $\zp$, coloron, or $\delta_6$, while one found at 5.2 TeV could be a $\Phi_6$, $\omega_3$ or coloron (or, with sufficient integrated luminosity, even a $\delta_6$ or a $\zp$).

In many situations where a dijet resonance of a given mass discovered at LHC could correspond to more than one class of particle, measuring $D_{col}$ will suffice to distinguish among them.  A leptophobic $\zp$ would not be confused with any of the weak-singlet scalar diquarks (except, possibly, the $\delta_6$ near the top of the mass range for a given integrated luminosity).  Nor would any of the color-sextet diquarks be mistaken for one another.  The color-triplet diquark and the coloron are, likewise, distinct by this measure.

In other cases, the measurement of $\dcol$ will suffice to show that a resonance is a diquark, and yet may not have sufficient precision to determine which kind of diquark state has been found.  For instance, at masses of order 6 TeV, the sextet states are all distinct from one another, but the $\omega_3$ overlaps both the $\Phi_6$ and $\phi_6$. Measuring the color flow~\cite{Maltoni:2002mq,Kilian:2012pz,Curtin:2012rm,Gallicchio:2010sw} in the events may be of value here.

In other cases, the measurement of $\dcol$ may leave us unsure as to whether a vector boson (coloron) or a color-sextet $\delta_6$ diquark has been discovered (e.g., at masses of order 3.5 TeV). In this case, measuring the angular distributions of the final state jets may assist in further distinguishing the possibilities \cite{Harris:2011bh}.

\section{Discussion}
\label{sec:discussions}

The current run of the LHC has the potential to discover a new dijet resonance, opening the doors to an era of physics beyond the standard model.  The simple topology and large production rate for a dijet final state will not only promote discovery, but also aid in the determination of crucial properties of the new resonance. Because the color discriminant variable~\cite{Atre:2013mja}, $\dcol$, is constructed from measurements available directly after the discovery of the resonance via the dijet channel, namely, its mass, its total decay width, and its dijet cross section, this variable can be valuable in identifying the nature of a newly discovered state ~\cite{Atre:2013mja,Chivukula:2014npa,Chivukula:2014pma}.

In this work, we have extended the color discriminant variable technique in two directions. First, we have placed the theoretical discussion of the variable in more general language that shows its broader applicability and its relationship to the properties of the partons involved in production and decay of the resonance.  Second, we have shown that $\dcol$ may be used both to identify scalar diquark resonances as color triplet or color sextet states and to distinguish them from color-neutral or color-octet vector bosons.

In addition to the dijet resonances that we have discussed in detail in this work, other exist -- and they are also amenable to analysis via the color discriminant variable. One example would be states whose decay products include gluons: scalar resonances decaying to $gg$ and excited quarks decaying to $qg$.  As shown in~\cite{Chivukula:2014pma}, $\dcol$ can distinguish moderate width diquarks ($qq$) and vector bosons ($q\bar{q}$), whose decay products do not include gluons, from resonances whose decays do include final state gluons.  When the resonance is too narrow for $\dcol$ to be measurable, studying the jet energy profile of the final state jets can be a valuable alternative.

We hope the color discriminant variable will called upon soon to identify a new dijet resonance discovered at the LHC.

\Acknowledgments
The authors thank Natascia Vignaroli for useful discussions. This material is based upon work supported by the National Science Foundation under Grant No. PHY-0854889. We wish to acknowledge the support of the Michigan State University High Performance Computing Center and the Institute for Cyber Enabled Research. PI is supported by Development and Promotion of Science and Technology Talents Project (DPST), Thailand.  EHS and RSC thank the Aspen Center for Physics and the NSF Grant \#1066293 for hospitality during the writing of this paper.



\begin{thebibliography}{99}
\bibitem{Chivukula:2015zma}
  R.~S.~Chivukula, P.~Ittisamai, K.~Mohan and E.~H.~Simmons,
  Phys.\ Rev.\ D {\bf 92}, no. 7, 075020 (2015)
  [arXiv:1507.06676 [hep-ph]].

\bibitem{ATLAS:2012pu}
  G.~Aad {\it et al.} [ATLAS Collaboration],
  JHEP {\bf 1301}, 029 (2013)
  [arXiv:1210.1718 [hep-ex]].

\bibitem{Chatrchyan:2013qha}
  S.~Chatrchyan {\it et al.} [CMS Collaboration],
  Phys.\ Rev.\ D {\bf 87}, no. 11, 114015 (2013)
  [arXiv:1302.4794 [hep-ex]].

\bibitem{CMS:kxa}
  [CMS Collaboration],
  CMS-PAS-EXO-12-059.

\bibitem{Atre:2013mja}
  A.~Atre, R.~S.~Chivukula, P.~Ittisamai and E.~H.~Simmons,
  Phys.\ Rev.\ D {\bf 88}, 055021 (2013)
  [arXiv:1306.4715 [hep-ph]].

\bibitem{Chivukula:2014npa}
  R.~Sekhar Chivukula, P.~Ittisamai and E.~H.~Simmons,
  Phys.\ Rev.\ D {\bf 91}, no. 5, 055021 (2015)
  [arXiv:1406.2003 [hep-ph]].

\bibitem{Chivukula:2014pma}
  R.~Sekhar Chivukula, E.~H.~Simmons and N.~Vignaroli,
  Phys.\ Rev.\ D {\bf 91}, no. 5, 055019 (2015)
  [arXiv:1412.3094 [hep-ph]].

\bibitem{Harris:2011bh}
  R.~M.~Harris and K.~Kousouris,
  Int.\ J.\ Mod.\ Phys.\ A {\bf 26}, 5005 (2011)
  [arXiv:1110.5302 [hep-ex]].

\bibitem{Han:2009ya}
  T.~Han, I.~Lewis and T.~McElmurry,
  JHEP {\bf 1001}, 123 (2010)
  [arXiv:0909.2666 [hep-ph]].

\bibitem{Gumus:2006mxa}
  K.~Gumus, N.~Akchurin, S.~Esen and R.~M.~Harris,
  CMS-NOTE-2006-070, CERN-CMS-NOTE-2006-070.

\bibitem{Maltoni:2002mq}
  F.~Maltoni, K.~Paul, T.~Stelzer and S.~Willenbrock,
  Phys.\ Rev.\ D {\bf 67}, 014026 (2003)
  [hep-ph/0209271].

\bibitem{Kilian:2012pz}
  W.~Kilian, T.~Ohl, J.~Reuter and C.~Speckner,
  JHEP {\bf 1210}, 022 (2012)
  [arXiv:1206.3700 [hep-ph]].

\bibitem{Curtin:2012rm}
  D.~Curtin, R.~Essig and B.~Shuve,
  Phys.\ Rev.\ D {\bf 88}, 034019 (2013)
  [arXiv:1210.5523 [hep-ph]].

\bibitem{Gallicchio:2010sw}
  J.~Gallicchio and M.~D.~Schwartz,
  Phys.\ Rev.\ Lett.\  {\bf 105}, 022001 (2010)
  [arXiv:1001.5027 [hep-ph]].


\end{thebibliography}
\end{document}